\documentclass[11pt]{article}
\usepackage{graphicx}
\usepackage{amssymb,amsmath,amsfonts}

\def\Re{{\cal R \mskip-4mu \lower.1ex \hbox{\it e}\,}}
\def\Im{{\cal I \mskip-5mu \lower.1ex \hbox{\it m}\,}}
\def\ie{{\it i.e.}}
\def\eg{{\it e.g.}}

\def\sub#1{_{\lower.25ex\hbox{$\scriptstyle#1$}}}
\def\tev{\,{\ifmmode\mathrm {TeV}\else TeV\fi}}
\def\gev{\,{\ifmmode\mathrm {GeV}\else GeV\fi}}
\def\mev{\,{\ifmmode\mathrm {MeV}\else MeV\fi}}
\def\mpl{\ifmmode M_{pl}\else $M_{pl}$\fi}
\def\mpl{\ifmmode \overline M_{Pl}\else $\bar M_{Pl}$\fi}
\def\to{\rightarrow}

\def\subw{_{\rm w}}
\def\mh{\ifmmode m\sbl H \else $m\sbl H$\fi}
\def\mch{\ifmmode m_{H^\pm} \else $m_{H^\pm}$\fi}
\def\mt{\ifmmode m_t\else $m_t$\fi}
\def\mc{\ifmmode m_c\else $m_c$\fi}
\def\mz{\ifmmode M_Z\else $M_Z$\fi}
\def\mw{\ifmmode M_W\else $M_W$\fi}
\def\mws{\ifmmode M_W^2 \else $M_W^2$\fi}
\def\mhs{\ifmmode m_H^2 \else $m_H^2$\fi}   
\def\mzs{\ifmmode M_Z^2 \else $M_Z^2$\fi}
\def\mts{\ifmmode m_t^2 \else $m_t^2$\fi}
\def\mcs{\ifmmode m_c^2 \else $m_c^2$\fi}
\def\mchs{\ifmmode m_{H^\pm}^2 \else $m_{H^\pm}^2$\fi}
\def\ztwo{\ifmmode Z_2\else $Z_2$\fi}
\def\zone{\ifmmode Z_1\else $Z_1$\fi}
\def\mtwo{\ifmmode M_2\else $M_2$\fi}
\def\mone{\ifmmode M_1\else $M_1$\fi}
\def\tb{\ifmmode \tan\beta \else $\tan\beta$\fi}
\def\xw{\ifmmode x\subw\else $x\subw$\fi}
\def\ch{\ifmmode H^\pm \else $H^\pm$\fi}
\def\lum{\ifmmode {\cal L}\else ${\cal L}$\fi}
\def\inpb{\,{\ifmmode {\mathrm {pb}}^{-1}\else ${\mathrm {pb}}^{-1}$\fi}}
\def\infb{\,{\ifmmode {\mathrm {fb}}^{-1}\else ${\mathrm {fb}}^{-1}$\fi}}
\def\epem{\ifmmode e^+e^-\else $e^+e^-$\fi}
\def\ppb{\ifmmode \bar pp\else $\bar pp$\fi}
\def\bsg{\ifmmode B\to X_s\gamma\else $B\to X_s\gamma$\fi}
\def\bsll{\ifmmode B\to X_s\ell^+\ell^-\else $B\to X_s\ell^+\ell^-$\fi}
\def\bstt{\ifmmode B\to X_s\tau^+\tau^-\else $B\to X_s\tau^+\tau^-$\fi}
\def\lamt{\ifmmode \tilde\lambda\else $\tilde\lambda$\fi}
\def\shat{\ifmmode \hat s\else $\hat s$\fi}
\def\that{\ifmmode \hat t\else $\hat t$\fi}
\def\uhat{\ifmmode \hat u\else $\hat u$\fi}

\newskip\zatskip \zatskip=0pt plus0pt minus0pt
\def\matth{\mathsurround=0pt}
\def\lsim{\mathrel{\mathpalette\atversim<}}

\def\atversim#1#2{\lower0.7ex\vbox{\baselineskip\zatskip\lineskip\zatskip
  \lineskiplimit 0pt\ialign{$\matth#1\hfil##\hfil$\crcr#2\crcr\sim\crcr}}}

\def\grtsim{\,\,\rlap{\raise 3pt\hbox{$>$}}{\lower 3pt\hbox{$\sim$}}\,\,}
\def\lsim{\,\,\rlap{\raise 3pt\hbox{$<$}}{\lower 3pt\hbox{$\sim$}}\,\,}

\renewcommand{\thefootnote}{\fnsymbol{footnote}}

\begin{document} \begin{titlepage}
\rightline{\vbox{\halign{&#\hfil\cr
&SLAC-PUB-12481\cr
}}}
\begin{center}
\thispagestyle{empty} \flushbottom { {
\Large\bf Searching for Lee-Wick Gauge Bosons at the LHC
\footnote{Work supported in part
by the Department of Energy, Contract DE-AC02-76SF00515}
\footnote{e-mail:
rizzo@slac.stanford.edu}}}
\medskip
\end{center}

\centerline{Thomas G. Rizzo}
\vspace{8pt} 
\centerline{\it Stanford Linear
Accelerator Center, 2575 Sand Hill Rd., Menlo Park, CA, 94025}

\vspace*{0.3cm}

\begin{abstract}
In an extension of the Standard Model(SM) based on the ideas of Lee and Wick, Grinstein, O'Connell and Wise have found an interesting  
way to remove the usual quadratically divergent contributions to the Higgs mass induced by radiative corrections. Phenomenologically, 
the model predicts the existence of Terascale, negative-norm copies of the usual SM fields with rather unique properties: ghost-like 
propagators and negative decay widths, but with otherwise SM-like couplings. The model is both unitary and causal on macroscopic scales. 
In this paper we examine whether or not such states with these unusual properties can be uniquely identified as such at the LHC. We find 
that in the extended strong and electroweak gauge boson sector of the model, which is the simplest one to analyze, such an identification 
can be rather difficult. Observation of heavy gluon-like resonances in the dijet channel offers the best hope for this identification. 
\end{abstract}



\renewcommand{\thefootnote}{\arabic{footnote}} \end{titlepage} 

%
%
%

\section{Introduction and Background}

The mechanism of electroweak symmetry breaking remains mysterious. How does one generate the masses for the gauge bosons and fermions of the Standard 
Model(SM) without encountering fine-tuning and naturalness issues and the associated hierarchy problem? Over the next few years the ATLAS and CMS experiments 
at the LHC should begin to probe for answers to these important questions with potentially surprising results. Meanwhile, it is important for us to examine as 
many scenarios as possible which address these problems in order to prepare ourselves for these long-awaited data. 

Recently, Grinstein, O'Connell and Wise(GOW){\cite {GOW}} have extended an old idea based on higher-derivative theories, due to Lee and Wick(LW){\cite {LW}}, 
to the SM context in a gauge invariant way which solves the hierarchy problem, eliminating the quadratic divergence of the Higgs boson mass 
that one usually encounters. The essential feature of the GOW construction is the introduction of negative-normed states into the usual SM Hilbert space, in 
particular, one new massive degree of freedom (or one vector-like pair in the fermion case) for each of the conventional SM particles. The contributions of these 
exotic new particles to the Higgs mass quadratic divergence then cancels the SM contribution, partner by partner, leaving only logarithmic terms. For example, 
in the gauge sector, the following new 
fields are introduced: an $SU(3)_c$ octet of `gluons', $g_{LW}$, with mass $M_3$, an $SU(2)_L$ isotriplet of weak bosons, $W_{LW}^{0,\pm}$, of mass $M_2$ and 
a heavy $U(1)_Y$ hypercharge field, $B_{LW}$, with mass $M_1$. The interactions of these new fields with each other and with the familiar ones of the SM are 
given in Ref.{\cite {GOW}}; due to naturalness arguments and the present direct{\cite {limit}} and indirect{\cite {LEPEWWG}} 
experimental constraints on the existence of such particles, it is anticipated that their masses must lie not too far above $\simeq 1$ TeV. The masses of the LW 
gauge bosons themselves are found to be radiatively stable.  

While the detailed structure of such a theory raises questions of unitarity, 
causality and vacuum stability (which have been at least partially addressed in Ref.{\cite {concern}}), the purpose of the present paper is to address a purely 
phenomenological issue. As long as such states are not too massive, since their interactions are very similar to those of their conventional SM counterparts, 
it is already clear that they will be produced {\it and} observed at the LHC based on the results of other existing analyses{\cite {tdr}}. The issue we want to 
address here is the uniqueness of their production signatures, \ie, can we tell at the LHC that we have indeed produced these negative-metric LW fields and 
not something else? Due to their rather strange and unusual properties, to be discussed below, it would {\it a priori} seem rather straightforward to give an 
affirmative answer to this question. However, contrary to such expectations as we will see, it will be not be so easy for the LHC to uniquely identify 
such states even in the most optimistically possible experimental situations where LW gauge boson resonances are produced and large statistical samples are available. 

The essential phenomenological features of these new states for our study are straightforward to summarize and are important to remember in the analysis that follows: 
($i$) the propagators and decay widths of LW particles have signs which are opposite to those of the familiar SM fields; ($ii$) the couplings of LW gauge 
fields to SM fermions are exactly those of the corresponding SM gauge fields. ($i$) and ($ii$) taken together imply, \eg, that in any process where a conventional 
and a LW gauge field are both exchanged between massless SM fermions the amplitude behaves as 
\begin{equation}
\sim {{i}\over {p^2-M_{SM}^2+iM_{SM}\Gamma_{SM}}}-{{i}\over {p^2-M_{LW}^2+iM_{LW}\Gamma_{LW}}}\,,
\end{equation}
apart from other overall factors. In particular $\Gamma_{LW}<0$ has exactly the same magnitude as would a heavy copy of the relevant SM gauge field; this is 
reminiscent of the Sequential SM(SSM){\cite {tasi}} or of flat, TeV-scale extra dimensions where fermions are 
confined to the origin of the fifth dimension (apart from an additional numerical factor{\cite {flat}} of $\sqrt 2$). ($iii$) The LW 
fermions, in the absence of Yukawa interactions, can only be pair produced via the exchange of either SM and/or their LW partner 
gauge fields. Since these LW fields are quite heavy with masses perhaps in excess of 1 TeV, this will make the LW fermions relatively difficult to examine in detail at 
the LHC. Once the Yukawa interactions are turned on, the single production of LW fermion fields becomes possible via the Higgs sector, at least in principle. 
These couplings are the same ones which are responsible for the finite lifetimes of the LW fermions allowing them to decay.   
For the first two generations, the Yukawas are sufficiently small as to render production cross sections quite tiny. For the third generation, though the 
Yukawa couplings are now larger the cross sections remain small for single production. The dominant mechanism in the case of the LW top partner 
will be $\bar t t$ production with one of 
the tops off-shell and splitting to a Higgs and an LW top. A strong {\it upper} bound on this cross section can be obtained by considering ordinary $\bar ttH$  
production which will be significantly larger since the LW top must be more more massive than ordinary SM top. One finds that for a Higgs mass of 120(200) GeV, this 
cross section is $\sim 500(100)$ fb. For the LW case the cross section will be significantly reduced due to the additional massive LW top in the final state.  
On the otherhand, the production of a pair of strongly interaction LW fermions with masses of $\sim 0.5(1)$ TeV will be on the order of $\sim 40(1)$ pb which is far 
larger. Thus LW fermion pair production will likely be the dominant mechanism to study these states.  
($iv$) There are no tree-level trilinear couplings between two SM gauge fields and one LW gauge field. This means, \eg, in the QCD sector of the model 
that there is no $ggg_{LW}$ interaction though a $gg_{LW}g_{LW}$ one does exist. This further implies that a $g_{LW}$ resonance cannot be made in $gg$ collisions. 
($v$) Although there is mixing between the SM and LW fields, it is generally 
highly suppressed by mixing angles which scale as $\theta \sim M_{SM}^2/M_{LW}^2$. This implies that, \eg, $W-W_{LW}$ mixing is $O(10^{-3})$ and we will ignore all 
such effects in the discussions that follow. (As usual this mixing may be most important in 
the top quark sector.) From this discussion it would appear that the cleanest, most easily accessible channel to discover and {\it identify} new LW states is in the 
single production of strong and electroweak gauge LW resonances via $q\bar q$ annihilation and to this we now turn.

\section{Analysis}

Our first task here is to consider the resonant 
production of the gauge states $W_{LW}^{\pm,0},B_{LW}$ and $g_{LW}$ at the LHC. At tree-level this is rather straightforward as, 
in the absence of mixing ($v$), their couplings are exactly the same as SM fields, apart from those important sign differences ($i$). We also know which channels 
to examine: $pp \to \ell^+ \ell^-+X$ for $W_{LW}^0,B_{LW}$, $pp\to \ell^{\pm}E_T^{miss}+X$ for $W_{LW}^{\pm}$ and $pp\to jj+X$ for $g_{LW}$. 

Let us begin with the production of $W_{LW}^{\pm}$ as in this case there 
is only one new state being produced, the coupling structure is purely chiral, and we need not concern ourselves with large QCD backgrounds. As a warm-up let us first 
remind ourselves of the production properties of a more typical $W'${\cite {me}} defining their couplings to the quarks and leptons as
\begin{equation}
{{g_{wk}}\over {2\sqrt 2}} V_{ff'}C^{\ell,q}_i\bar f \gamma_\mu (1-h_i\gamma_5)f'W^\mu_i+ h.c.\,,
\end{equation}
where $g_{wk}$ is the usual weak coupling, $V$ is the conventional quark or neutrino mixing matrix, $C^{\ell,q}_i$ give the strength of the quark and leptonic 
couplings of the charged gauge boson, $W_i$, with helicity $h_i$. For the SM $W$, $C^{\ell,q},h=1$.   
Following the notation given in{\cite {me}}, the inclusive $pp\to W^+_i \to\ell^+ \nu+X$ differential cross section can then be written as 
{\footnote {Note that analogous expressions can also be written in the case of $W^-_i$ exchange by taking  $z\to -z$ and interchanging initial state quarks 
and anti-quarks in what appears below.}}
\begin{equation}
{{d\sigma}\over {d\tau~dy~dz}}= K~{{G_F^2M_W^4}\over {48\pi}} \sum_{qq'} |V_{qq'}|^2~\Big[SG_{qq'}^+(1+z^2)+2AG_{qq'}^-z\Big]\,,
\end{equation}
where $K$ is an approximate numerical factor that accounts for NLO and NNLO QCD corrections{\cite {nnlo}} which is roughly of order $\simeq 1.3$, $\tau=M^2/s$, 
with $M^2$ being the lepton pair invariant mass and $z=\cos \theta^*$, the center of mass frame scattering angle. Here,    
\begin{eqnarray}
S &=& \sum_{ij}P_{ij}(C_iC_j)^\ell (C_iC_j)^q(1+h_ih_j)^2\\ \nonumber
A &=& \sum_{ij}P_{ij}(C_iC_j)^\ell (C_iC_j)^q(h_i+h_j)^2\,, 
\end{eqnarray}
where the sums extend over all of the exchanged $W_i$ in the $s$-channel. We employ the notation  
\begin{equation}
P_{ij}=\shat{{(\shat-M_i^2)(\shat-M_j^2)+\Gamma_i\Gamma_jM_iM_j}\over {[(\shat-M_i^2)^2+\Gamma_i^2M_i^2][i\to j]}}\,,
\end{equation}
with $\shat=M^2$ being the square of the total collision energy and $\Gamma_i$ the total widths of the 
exchanged $W_i$ particles; note that $P_{ij}$ is symmetric in its indices. Furthermore, the following combinations of parton distribution functions appear: 
\begin{equation}
G^\pm_{qq'}=\Big[q(x_a,M^2)\bar q'(x_b,M^2)\pm q(x_b,M^2)\bar q'(x_a,M^2)]\,,
\end{equation}
where $q(q')$ is a $u(d)-$type quark and $x_{a,b}=\sqrt \tau e^{\pm y}$ are the corresponding parton density functions(PDFs). 
In most cases one usually converts the distribution over $z$ above into one over the transverse mass, $M_T$, formed from the final 
state lepton and the missing transverse energy associated with the neutrino; at fixed $M$, one has $z=(1-M_T^2/M^2)^{1/2}$. The 
resulting transverse mass distribution can then be written as 
\begin{equation}
{{d\sigma}\over {dM_T}} =\int _{M_T^2/s}^1~d\tau \int^Y_{-Y} ~dy~J(z\to M_T)~{{d\sigma}\over {d\tau~dy~dz}}\,,
\end{equation}
where $Y=min(y_{cut},-1/2 \log \tau)$ allows for a rapidity cut on the outgoing leptons and $J(z\to M_T)$ is the appropriate 
Jacobian factor{\cite {BP}}. In practice, $y_{cut}\simeq 2.5$ for the two LHC detectors.   
Note that ${{d\sigma}\over {dM_T}}$ will only pick out the $z$-even part of ${{d\sigma}\over {d\tau~dy~dz}}$ as well as the even combination of 
terms in the product of the parton densities, $G^+_{qq'}$ so that for our discussion it is sufficient to focus solely on the quantity $S$. 

So far we have been quite general but have had in mind a more conventional $W'$ that one usually encounters; what happens in the case of $W_{LW}$? As in the 
SM and SSM cases, one finds that $C^{\ell,q}_{LW},h_{LW}=1$ but here  
$\Gamma(W_{LW})=-\Gamma(W_{SSM})$ is {\it negative}. Furthermore, due to the new relative minus sign between the the $W$ and $W_{LW}$ propagators, the first, 
width-independent term in the numerator of $P_{SM,LW}$ is of {\it opposite} sign to that found in $P_{SM,SSM}$. It also important to notice that 
$P_{SSM,SSM}=P_{LW,LW}$; this tells us that measurements made near $M_T\simeq M$ or calculations based on the Narrow Width Approximation  
can {\it never} distinguish these two cases. However, in the $W-W_{LW}$ interference region, we are assured that the LW model cross section will differ from that 
obtained in the SSM case due to the additional relative minus signs.  

\begin{figure}[htbp]
\centerline{
\includegraphics[width=7.5cm,angle=90]{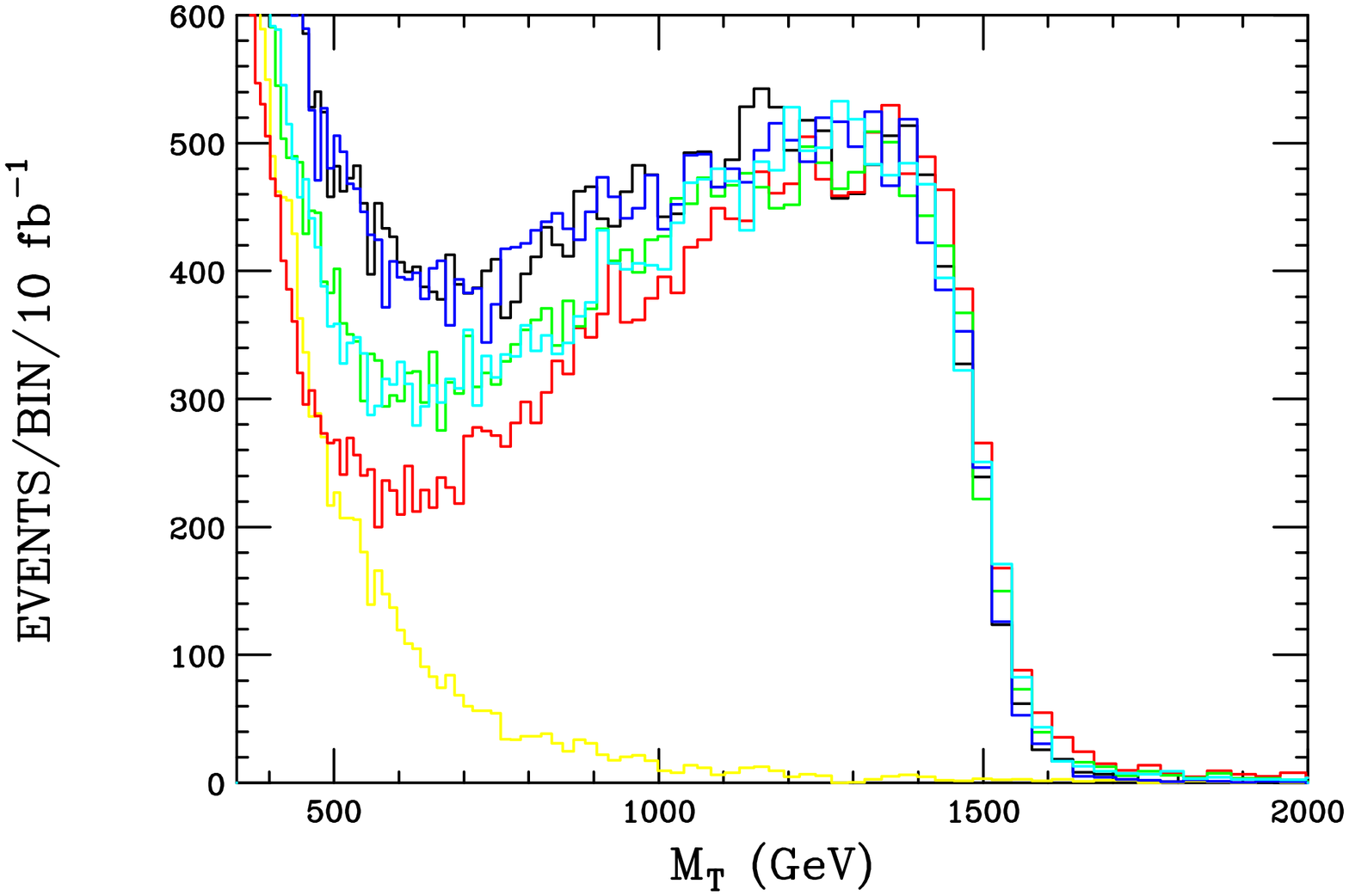}}
\vspace*{0.1cm}
\centerline{
\includegraphics[width=7.5cm,angle=90]{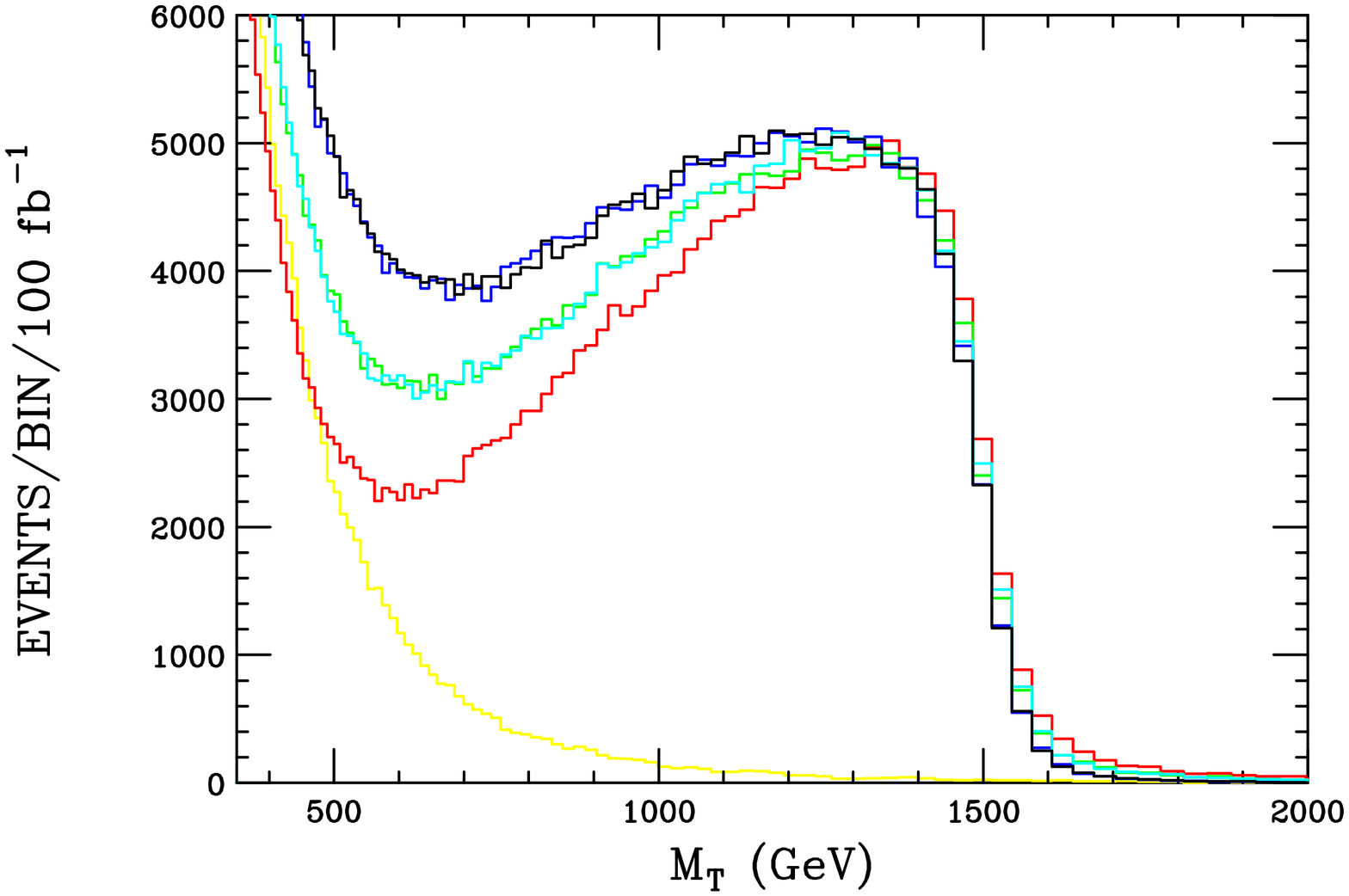}}
\vspace*{0.1cm}
\caption{Transverse mass distributions for $W'$ in various models at the LHC assuming a $W'$ mass of 1.5 TeV as described in the text. The upper(lower) panel assumes 
a luminosity of 10(100) $fb^{-1}$; the yellow histogram in both cases is the anticipated SM background. A cut of $|\eta_\ell| <2.5$ has been applied and the 
distribution has been smeared by $\delta M_T/M_T=2\%$ to simulate the ATLAS resolution for the electron final state.}
\label{fig1}
\end{figure}

Let us now compare the transverse mass distribution produced by $W_{LW}$ with several more conventional $W'$ models; to be specific and to have as much statistical 
power as possible we will assume that all the $W'$ have masses of 1.5 TeV. The results of these calculations are shown in Fig.~\ref{fig1} where we immediately see 
that there will be no difficulty in observing the $W'$ at the LHC for any of the cases considered. In this Figure, the red(green) histograms correspond to 
conventional $W'$ models 
with $C^{\ell,q}_{W'}=1,h_{W'}=1(-1)$ while the blue(cyan) histograms correspond to cases with $C^{\ell}_{W'}=-C^q_{W'}=1,h_{W'}=1(-1)$; the black histogram 
is the prediction of the LW scenario considered here. At first glance we see that indeed the LW model is clearly distinguishable from the SSM case as expected as 
well as from the Left Right Symmetric Model{\cite {LRM}}, which has  $C^{\ell,q}_{W'}=1,h_{W'}=-1$; it also differs from the $C^{\ell}_{W'}=-C^q_{W'}=1,h_{W'}=-1$ 
scenario which may occur in some models with extra dimensions. However, the prediction of the LW model and that for the case of 
$C^{\ell}_{W'}=-C^q_{W'}=1,h_{W'}=1$ are apparently indistinguishable. The two models 
{\it do} differ in their predictions algebraically: the term proportional to $M_W\Gamma_WM_{W'}\Gamma_{W'}$ in the numerator of $P_{WW'}$, when weighted with 
the relevant coupling factors, differs in sign in these two cases. This can be traced back to the negative decay width of $W_{LW}$. However the size of this 
particular term is quite small in comparison to all others due to the relatively narrow widths of the $W$ and $W_{LW}$ and is essentially impossible to dig out from 
the $M_T$ distribution. 

Is such a conspiratorial model realistic? Consider a 5D, $S^1/Z_2$ model of flat, TeV-scale extra dimensions{\cite {flat}} 
taking $R^{-1}\simeq 1.5$ TeV with leptons located at the origin of the extra dimension, $y=0$, and the quarks localized at $y=\pi R$. Such a model 
has been suggested to avoid proton decay constraints{\cite {nima}}. Then the lowest $W$ Kaluza-Klein(KK) excitation, due to its 5D wavefunction $\sim \cos y/R$, 
is found to have $C^{\ell}_{W'}=-C^q_{W'}={\sqrt 2},h_{W'}=1$, which is not very different from the toy model above. Brane kinetic terms{\cite {brane}} at 
$y=0,\pi R$ can easily reduce the $\sqrt {2} \to \simeq 1$ producing the above toy model while simultaneously reducing the couplings of the second and 
higher $W$ KK states making them difficult to observe. Thus it seems that with not much effort one can construct an example of a semi-realistic model which is difficult 
to distinguish from the LW one considered here, at least in this particular channel.  

\begin{figure}[htbp]
\centerline{
\includegraphics[width=7.5cm,angle=90]{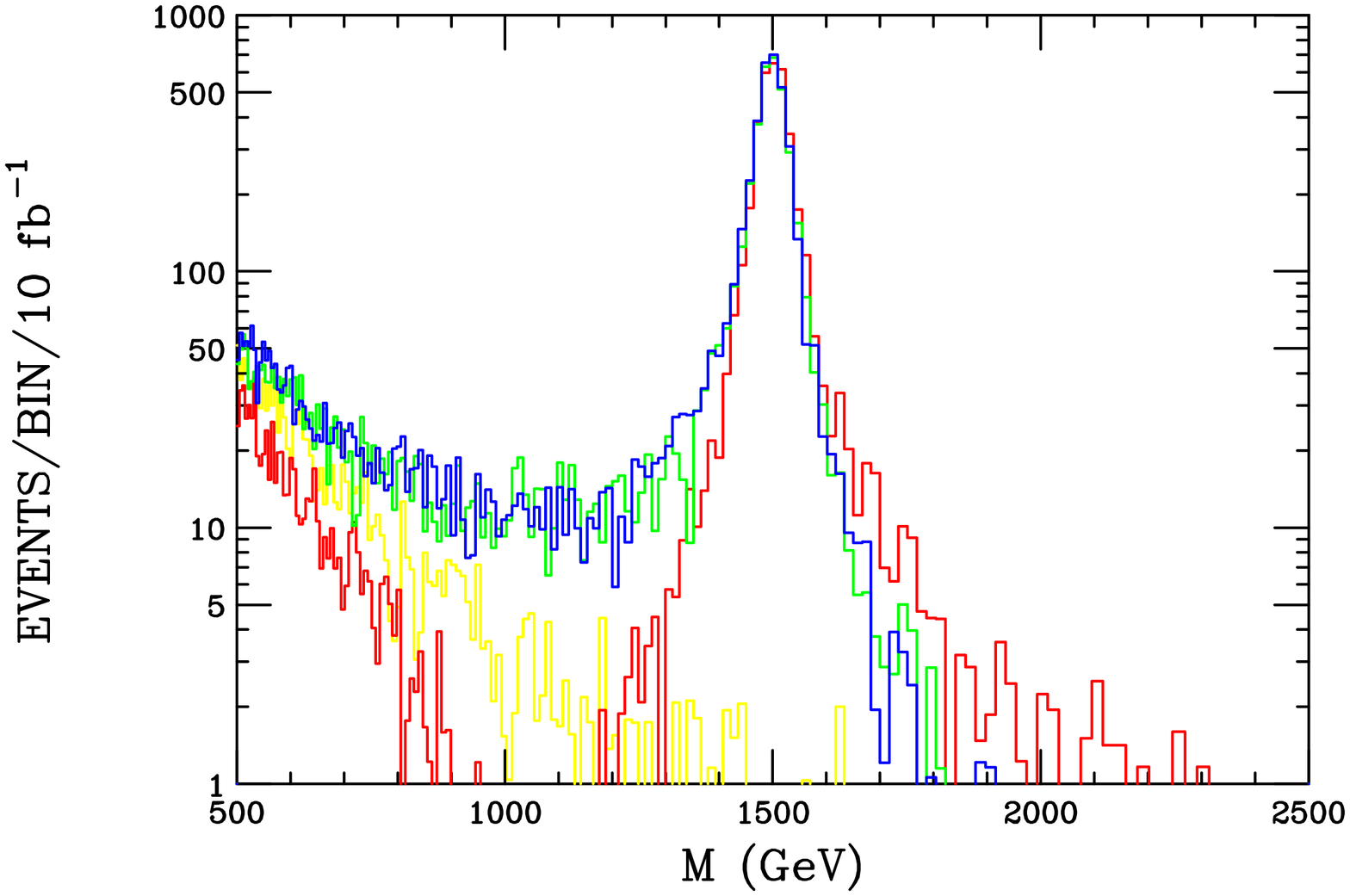}}
\vspace*{0.1cm}
\centerline{
\includegraphics[width=7.5cm,angle=90]{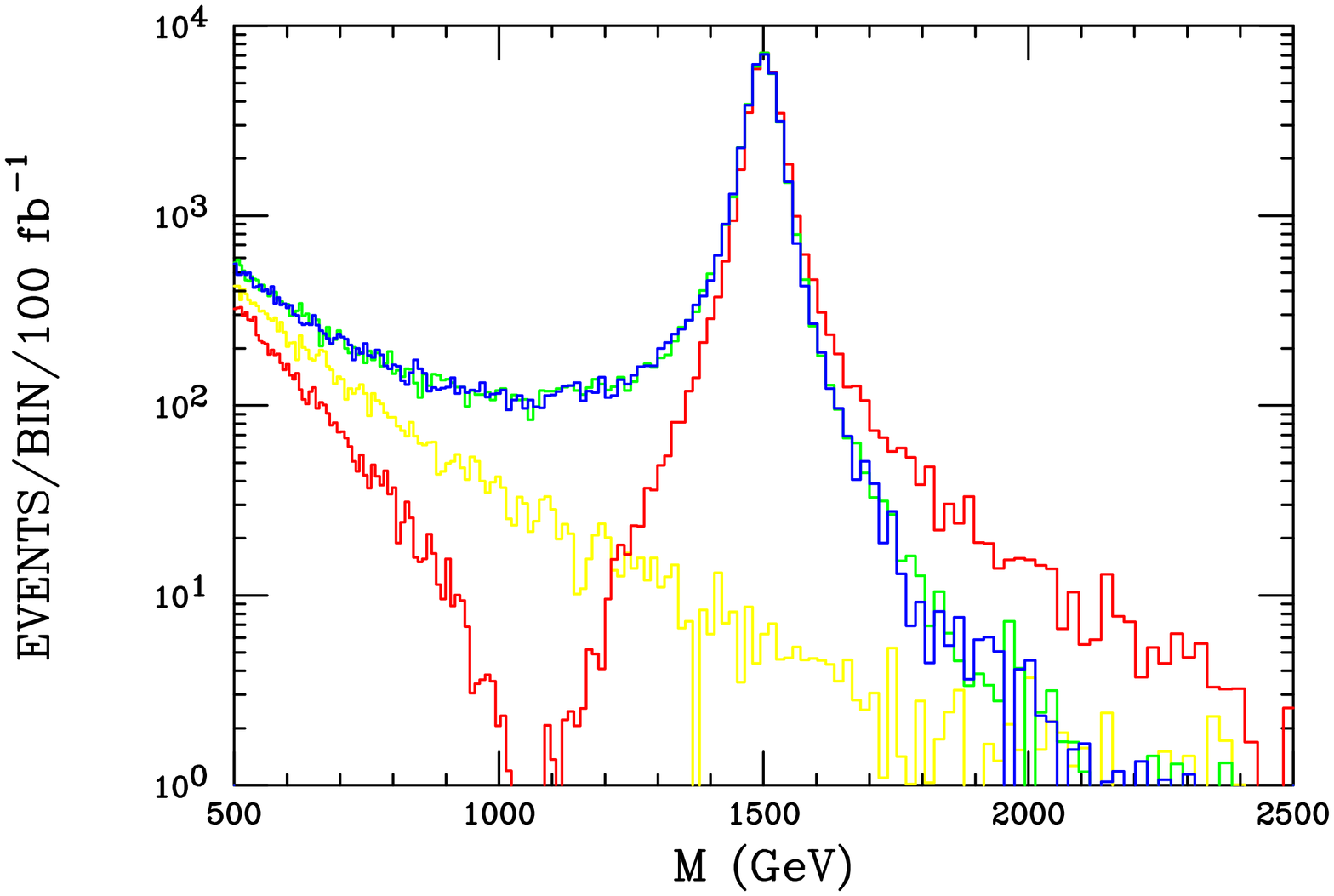}}
\vspace*{0.1cm}
\caption{Dilepton pair mass distributions at the LHC for the two models as described in the text. The upper(lower) panel assumes a luminosity of 10(100) 
$fb^{-1}$; the yellow histogram in both cases is the anticipated SM background. A cut of $|\eta_\ell| <2.5$ has been applied to both final state leptons and the 
mass distribution has been smeared by $\delta M/M=1\%$ to simulate the ATLAS resolution for the electron final state.}
\label{fig2}
\end{figure}

At this point the reader may say that the reason that the LW and $C^{\ell}_{W'}=-C^q_{W'}=1,h_{W'}=1$ models are apparently indistinguishable is due to the 
presence of the smeared out Jacobian peak structure in the $M_T$ distribution instead of a true dilepton mass peak which might be much cleaner. To that end we turn 
to the case of $W_{LW}^0,B_{LW}$ production in dilepton pairs; in order to reduce the number of parameters and to keep the statistics high we will assume 
that these two states are degenerate with a mass of 1.5 TeV. Clearly, the situation would be far more difficult to analyze if this were not the case. If the 
$W_{LW}^0$ and $B_{LW}$ resonances occurred at two random mass values there would be little chance of the data highlighting any unique LW features. 
Given our $W'$ experience, we know that there will be no issue of LW/SSM distinquishability. However, we should pay special attention to a KK-like modified version 
of the SSM where a heavy copy of the photon also exists and where the quark couplings have their signs reversed. Based on the previous $W'$ analysis we expect that 
this will be the case where model distinction will be most difficult. This follows by direct analogy to what occurs 
in the conspiratorial $W'$ toy model as well as in the TeV-scale 
extra dimensional KK scenario described above{\footnote {It is also known{\cite {old}} that degenerate $Z/\gamma$ KK states and conventional $Z'$ excitations are 
distinguishable at the LHC so we need not concern ourselves with this type of ambiguity.}}. 
The cross section calculation in this channel proceeds as in the $W'$ case above 
with only some minor modifications so we give no details here; Fig.~\ref{fig2} shows the results 
of this analysis. Here the red histogram corresponds to production of degenerate, conventional heavy copies of the $Z$ and $\gamma$; the same is true for 
the blue histogram case except now the sign of the quark couplings are reversed in comparison to the SM.  The green histogram is the result predicted 
by the LW scenario under study. 
Unfortunately, we again see that as long as we are able to have quarks and leptons with opposite overall signs in their otherwise SM couplings we cannot 
uniquely identify the LW model; the only difference continues to arise solely in the width-dependent interference terms and they are just too small in magnitude to 
make any appreciable modification to the cross section even in this somewhat cleaner final state. Thus, it would seem that as long as we 
have initial and final states which are fermions of different types, \ie, quarks and leptons, we can always play this same game and hide the uniqueness of the LW 
gauge fields by suitably adjusting the relative signs of these couplings.

\begin{figure}[htbp]
\centerline{
\includegraphics[width=7.5cm,angle=90]{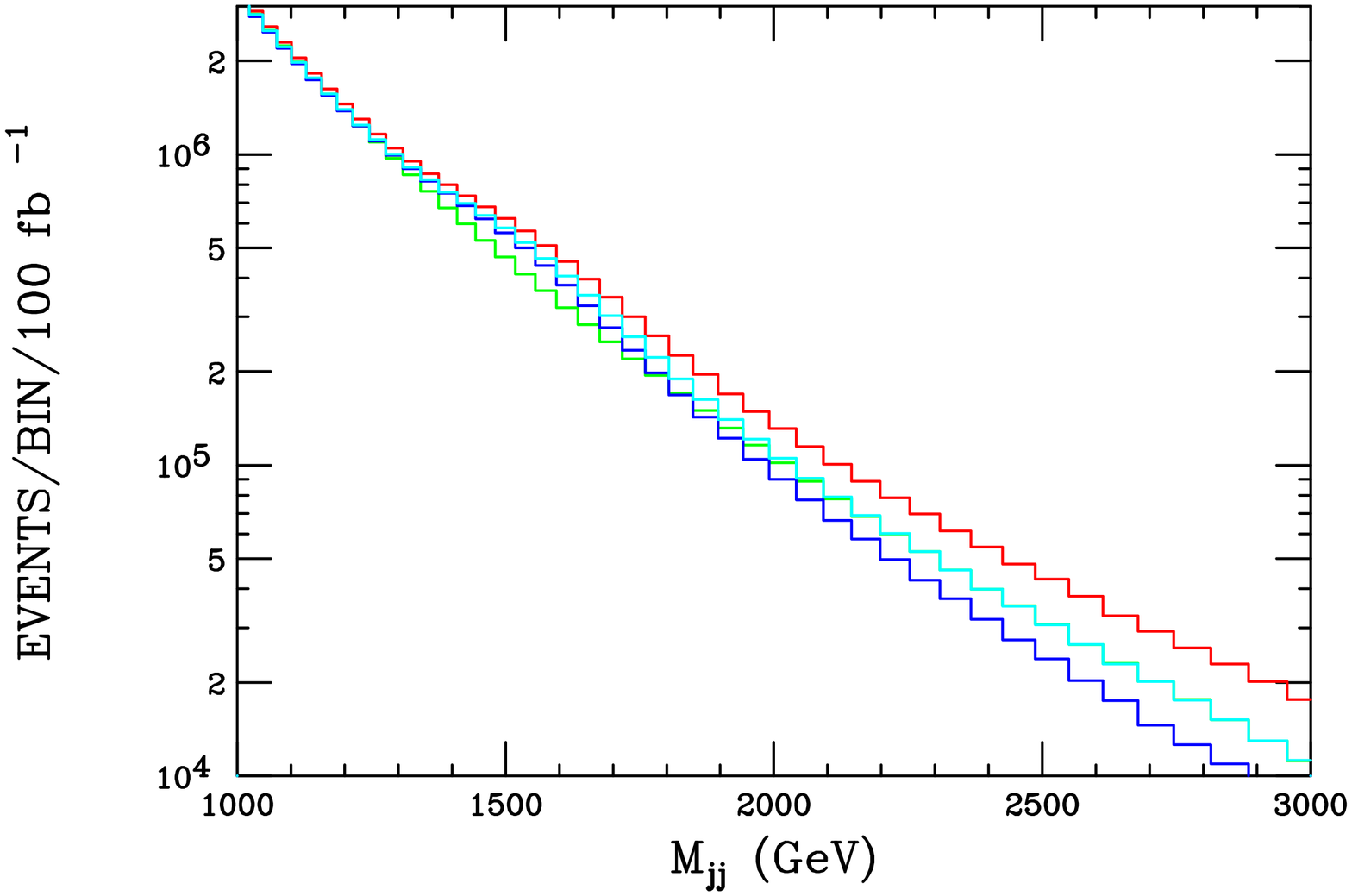}}
\vspace*{0.1cm}
\centerline{
\includegraphics[width=7.5cm,angle=90]{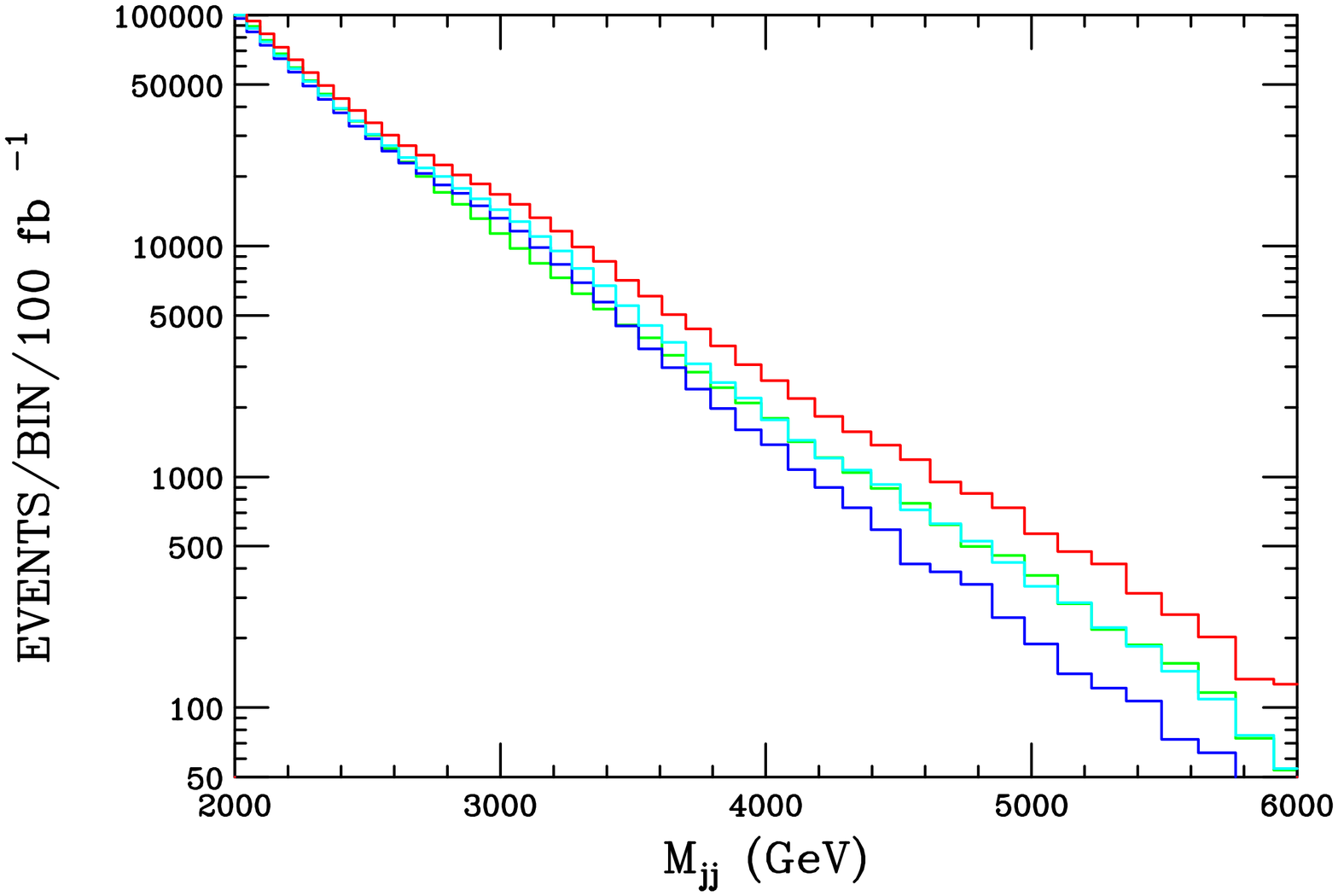}}
\vspace*{0.1cm}
\caption{Dijet production at the LHC as a function of the jet pair mass at high integrated luminosities assuming a resonance mass of 1.5(3) TeV in the top(bottom) 
panel. A pair of cuts, $|\eta_j|<1$ and $p_T>0.3M_{jj}$, have been applied and the distribution has 
been smeared by a mass resolution of  $\delta M_{jj}/M_{jj}=8\%$ to approximately match the CMS search analysis in this mass regime{\cite {rob}}. The green 
histogram is the SM QCD LO ($\mu=p_T$) background while the red(blue) histogram is for a heavy KK-like copy of the SM gluon($g_{LW}$) with SM couplings. The 
cyan histogram is the corresponding result for axigluons.}
\label{fig3}
\end{figure}

\begin{figure}[htbp]
\centerline{
\includegraphics[width=7.5cm,angle=90]{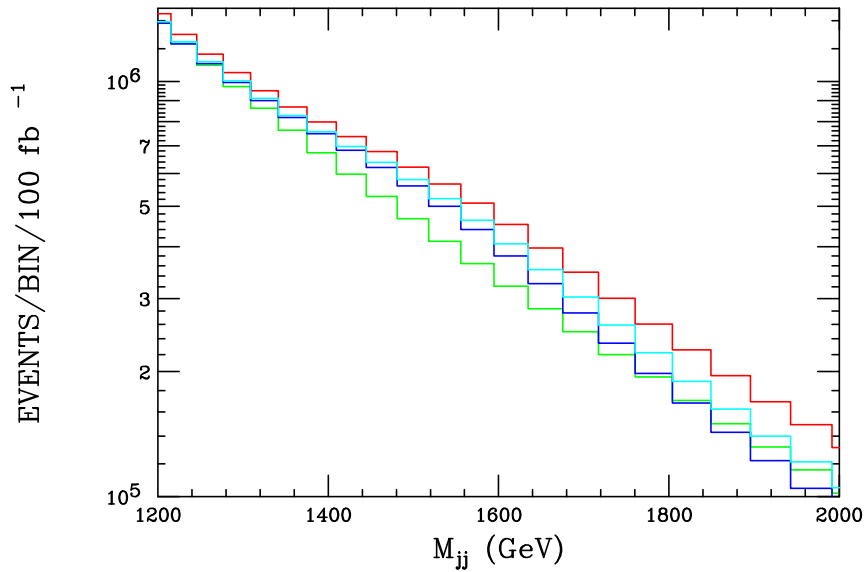}}
\vspace*{0.1cm}
\caption{Same as the top panel of the previous figure but now with a magnified view of the resonant region.}
\label{fig4}
\end{figure}

The most obvious way to circumvent this issue at the LHC is to turn to a case where the initial and final state particles are of the {\it same} type; this is exactly 
what happens in the 
QCD sector where we can search for resonances in the dijet channel{\footnote {We can also study the neutral gauge boson sector above in Bhabha scattering at a high 
energy $e^+e^-$ collider.}}. Here the only two models with QCD strength, flavor-independent spin-1 exchanges to compare 
are that of a heavy KK-like{\cite {flat}} gluon with SM couplings and the present $g_{LW}$ scenario. There is also, of course, the axigluon model{\cite {axigluon}}, 
but it does not interfere with SM gluon exchange in the resonant channel when the squared matrix element is symmetrically 
integrated over the jet production angles; however, we will include this possibility for completeness as it leads to a distinctive dijet distribution. 
In all these cases, $ggg'$-type couplings are 
absent, ($iv$) above, so that only quark and anti-quark initial states contribute to the cross section of the resonating state. The results of these considerations 
are shown in Fig.~\ref{fig3} again taking an excitation mass of 1.5 TeV. (All calculations were performed in LO at the $2\to 2$ parton level assuming a scale $\mu=p_T$.)
Note that we have applied several cuts to increase the effect of the resonance contribution as was done by the CMS collaboration in their dijet resonance search 
analysis{\cite {rob}}. Unfortunately, though in all model cases an excess is observed in the resonant region, there is no strong peaking evident due to the large 
non-resonant SM backgrounds and the effects of the jet pair 
mass resolution smearing. The predictions of the three models are clearly distinct and they all will lead so a significant deviation from ordinary QCD expectations. 
As expected, the axigluon prediction differs from QCD only in the resonant region itself while the other two models also display distinctive deviations from the SM 
in the behavior of the long tails at higher dijet masses. This is best seen by comparing Fig.~\ref{fig3} and Fig.~\ref{fig4} which shows an 
expanded view of the resonance region. In particular, in the LW case we see a destructive interference with the SM as might be expected 
at large masses due to the 
opposite sign in the propagator. On the otherhand, the heavy gluon scenario leads to constructive interference at these same masses. This pattern persists in the 
case where the new state is somewhat heavier in a statistically significant way as can be seen in the lower panel of Fig.~\ref{fig3}. 

The real difficulty with these results is that we have relied on a LO, parton-level analysis and it is not clear how well these visible differences will persist when 
NLO corrections and detector effects are included. Furthermore, the uncertainty in the PDFs is greatest at large dijet invariant masses which leads to 
some uncertainty in the predictions of SM QCD{\footnote {In the present analysis we make use of the CTEQ6M PDFS{\cite {CTEQ}}.}}. 
However it is likely that the qualitative result, that the LW(axigluon,KK) model prediction is 
below(equal to,above) the SM background at invariant masses significantly above the resonance region, still persists.  
To address these issues a more detailed analysis is clearly required. In either case it appears that the dijet channel holds out the greatest promise for uniquely 
identifying the LW model.

\section{Discussion and Conclusions}

In this paper we have examined whether or not the exotic, negative-normed gauge states predicted by the GOW model can be uniquely identified as such at the LHC. We 
have clearly demonstrated that the exotic properties of the weakly interacting states, which are produced in the Drell-Yan channel, could easily be faked by more 
conventional models, \ie, those based on flat TeV-scale extra dimensions with differently localized quarks and leptons. The dijet channel in the case of the 
analog heavy LW gluon was found to offer us a greater likelihood for distinguishing models even if the masses of these new states are rather high as the three models 
we considered led to qualitatively different behavior. To insure that this is the case requires a NLO calculation and a more serious analysis of detector effects.

\noindent{\Large\bf Acknowledgments}

The author would like to thank T.D. Lee and G.C. Wick for teaching him the elements of the Standard Model and Quantum Field Theory as a graduate student many years 
ago. He would also like to thank R. Harnik, J.L. Hewett and R. Kitano for discussions related to this work.

%
\def\MPL #1 #2 #3 {Mod. Phys. Lett. {\bf#1},\ #2 (#3)}
\def\NPB #1 #2 #3 {Nucl. Phys. {\bf#1},\ #2 (#3)}
\def\PLB #1 #2 #3 {Phys. Lett. {\bf#1},\ #2 (#3)}
\def\PR #1 #2 #3 {Phys. Rep. {\bf#1},\ #2 (#3)}
\def\PRD #1 #2 #3 {Phys. Rev. {\bf#1},\ #2 (#3)}
\def\PRL #1 #2 #3 {Phys. Rev. Lett. {\bf#1},\ #2 (#3)}
\def\RMP #1 #2 #3 {Rev. Mod. Phys. {\bf#1},\ #2 (#3)}
\def\NIM #1 #2 #3 {Nuc. Inst. Meth. {\bf#1},\ #2 (#3)}
\def\ZPC #1 #2 #3 {Z. Phys. {\bf#1},\ #2 (#3)}
\def\EJPC #1 #2 #3 {E. Phys. J. {\bf#1},\ #2 (#3)}
\def\IJMP #1 #2 #3 {Int. J. Mod. Phys. {\bf#1},\ #2 (#3)}
\def\JHEP #1 #2 #3 {J. High En. Phys. {\bf#1},\ #2 (#3)}

\end{document}